%% file: main.tex
\documentclass[sigconf]{acmart}

\copyrightyear{2017} 
\acmYear{2017} 
\setcopyright{acmcopyright}
\acmConference{CCS '17}{October 30-November 3, 2017}{Dallas, TX,
USA}\acmPrice{15.00}\acmDOI{10.1145/3133956.3134038}
\acmISBN{978-1-4503-4946-8/17/10}

\usepackage{booktabs} 
\usepackage{multirow}
\usepackage{listings}
\usepackage{makecell}
\usepackage{xspace}
\usepackage{balance}

\newcommand{\ignore}[1]{}
\newcommand{\revised}[1]{}
\newcommand\revise[1]{#1}

\newtheorem{nonumbertheorem}{Theorem}
\newtheorem{attack}[nonumbertheorem]{Vector}

\newcommand\yinqian[1]{\textcolor{blue}{\{\textbf{yinqian:} {\em#1}\}}}

\newcolumntype{C}[1]{>{\centering\let\newline\\\arraybackslash\hspace{0pt}}m{#1}}
\newcolumntype{L}[1]{>{\raggedright\let\newline\\\arraybackslash\hspace{0pt}}m{#1}}
\newcommand{\etal}{{\textit{et al.} }}

\newcounter{packednmbr}

\newenvironment{packeditemize}{
\begin{list}{$\bullet$}{
\setlength{\labelwidth}{8pt}
\setlength{\itemsep}{0pt}
\setlength{\leftmargin}{\labelwidth}
\addtolength{\leftmargin}{\labelsep}
\setlength{\parindent}{0pt}
\setlength{\listparindent}{\parindent}
\setlength{\parsep}{0pt}
\setlength{\topsep}{3pt}}}{\end{list}}

\newcommand{\dejavu}{{D{\small\'{E}J\`{A}} V{\small{U}}}}

\newcommand{\accessed}{\textit{accessed} flags\xspace}

\newcommand{\dirtyflags}{\textit{dirty} flags\xspace}

\newcommand{\flushflush}{\textsc{Flush+Flush}\xspace}
\newcommand{\flushreload}{\textsc{Flush+Reload}\xspace}
\newcommand{\primeprobe}{\textsc{Prime+Probe}\xspace}

\begin{document}

\title[Leaky Cauldron on the Dark Land]{Leaky Cauldron on the Dark Land: Understanding Memory Side-Channel Hazards in SGX}  
\author{Wenhao Wang$^{1}$, Guoxing Chen$^{3}$, Xiaorui Pan$^2$, Yinqian Zhang$^3$, XiaoFeng Wang$^2$, \\Vincent Bindschaedler$^4$, Haixu Tang$^2$, Carl A. Gunter$^4$}
\authornote{Work was done when the first author was at Indiana University Bloomington. \\Email: \{ww31, xiaopan, xw7, hatang\}@indiana.edu, \{chenguo, yinqian\}@cse.ohio-state.edu, \{bindsch2, cgunter\}@illinois.edu.}
\affiliation{
        \institution{$^1$SKLOIS, Institute of Information Engineering, Chinese Academy of Sciences \& Indiana University Bloomington \\
        $^2$Indiana University Bloomington\\$^3$The Ohio State University \\$^4$University of Illinois at Urbana-Champaign}
}


\renewcommand{\shortauthors}{W. Wang et al.}

\begin{abstract}

Side-channel risks of Intel SGX have recently attracted great attention. Under
the spotlight is the newly discovered page-fault attack, in which an OS-level
adversary induces page faults to observe the page-level access patterns of a
protected process running in an SGX enclave. With almost all proposed defense
focusing on this attack, little is known about whether such efforts indeed
raise the bar for the adversary, whether a simple variation of the attack
renders all protection ineffective, not to mention an in-depth understanding of
other attack surfaces in the SGX system. In the paper, we report the first step
toward systematic analyses of side-channel threats that SGX faces, focusing
on the risks associated with its memory management. Our research identifies 8
potential attack vectors, ranging from TLB to DRAM modules. More importantly, we highlight the common misunderstandings about SGX memory side channels, demonstrating that high frequent AEXs can be avoided when recovering EdDSA secret key through a new page channel and fine-grained monitoring of enclave programs (at the level of 64B) can be done through combining both cache and cross-enclave DRAM channels. Our findings reveal the gap between the ongoing security research on SGX and its side-channel weaknesses, redefine the side-channel threat model for secure enclaves, and can provoke a discussion on when to use such a system and how to use it securely. 

\end{abstract}

\maketitle

\ignore{Particularly, SPM utilizes the \accessed on page table entries, timing and
HyperThreading to infer sensitive operations of an enclave process.  The attack
avoids producing a large number of interrupts and even works on the program
running on a single page, at an exceedingly low cost compared to the page-fault
attack. We also show that a novel enclave-based DRAM attack can defeat all
proposed defense and can even achieve the precision of the \flushreload attack
when combined with the cache side channels.}

\input{tex/intro.tex}

\input{tex/background.tex}

\input{tex/surface.tex}

\input{tex/attack.tex}
\input{tex/discussion.tex}

\input{tex/relatedwork.tex}
\input{tex/conclusion.tex}
\input{tex/acknowledgement.tex}



\balance
\bibliographystyle{ACM-Reference-Format}
\bibliography{paper}


\end{document}

%% file: tex/intro.tex
\section{Introduction}
\label{sec:introduction}

A grand security challenge today is how to establish a trusted execution
environment (TEE) capable of protecting large-scale, data-intensive computing.
This is critical for the purposes such as outsourcing analysis of sensitive data
(e.g., electronic health records) to an untrusted cloud. Serving such purposes
cannot solely rely on cryptographic means, for example, fully homomorphic
encryption, which is still far too slow to handle the computing task of a
practical scale. A promising alternative is made possible by recent hardware progress
such as Intel Software Guard Extension (SGX)~\cite{anati2013innovative}.
SGX offers protection for data and code with a secure
\textit{enclave} designed to be resilient to attacks from its host operating
system or even system administrators. Such protection is offered as a feature of
Intel's mainstream CPUs (i.e., Skylake and Kaby Lake), and characterized by its small
trusted computing base (TCB), including just the CPU package, and the potential
to scale its performance with the capability of the processors. However, the
simplicity of the design forces an enclave program to utilize resources (memory,
I/O, etc.) partially or fully controlled by the untrusted OS, and therefore
potentially subjects it to \textit{side-channel} attacks, in which the adversary
outside the enclave could \textit{infer} sensitive information inside from
observed operations on the shared resources.


\vspace{3pt}\noindent\textbf{SGX side-channel hazards}. Unfortunately, the threat has been found to be more realistic and serious than thought: prior studies have shown that an adversary with a full control of the OS can manipulate the page tables of the
code running in the enclave-mode---a CPU mode protected by SGX---to induce page
faults during its execution; through monitoring the occurrences of the faults in
a relatively noise-free environment created by the adversary, he could identify
the program's execution trace at the page level, which is shown to be sufficient
for extracting text documents, outlines of images from popular application
libraries~\cite{Xu:2015:controlled} and compromising cryptographic
operations~\cite{Shinde:2015:PYF}. In our paper, we call these attacks the page-fault side-channel attacks.

Intel's position on these side-channel attacks offers much food for thought. They admit that SGX does not defend against four side-channel attack vectors: power statistics, cache miss statistics, branch timing and page accesses via page tables~\cite{intelsgx}. Facing the security threats due to these side channels, Intel recommends that ``it would be up to the independent software vendors to design the enclave in a way that prevents the leaking of side-channel information.~\cite{enclaveguide}'', though they actually work actively with academia and open source partners to help mitigate the threats\ignore{ on the software layers}~\cite{sgxsidechannel}. 

It is clear, from Intel's statements, radical hardware changes to address these side-channel problems\ignore{ that require changes to the SGX's hardware design} (e.g., defeating page-fault side channels by keeping the entire page tables inside the enclave~\cite{Costan:2016:sanctum}) are unlikely to happen. As a result, software vendors are left with the daunting tasks of understanding the security impacts of the SGX side channels and developing practical techniques to mitigate the threats when building their SGX applications. 


Given the importance of the problem, recent years have already witnessed mushrooming of the attempts to address  SGX side channel threats~\cite{Shinde:2015:PYF,shih:tsgx,Chen:2017:dejavu}, for example, by placing sensitive code and data within the same page~\cite{Shinde:2015:PYF}, or by detecting page faults through hardware supports~\cite{shih:tsgx} or timed-execution~\cite{Chen:2017:dejavu}. However, these studies were primarily targeting the page-fault attacks. Although the concern about this demonstrated attack vector is certainly justified, the sole attention on the page-fault attack can be inadequate. After all,
tasking a safe heaven built upon minimum software support with complicated
computing missions (e.g., data-intensive computing) can potentially open many
avenues for inside information to trickle out during the interactions with the
outside, when the external help is needed to accomplish the missions. Focusing on memory
alone, we can see not only a program's virtual memory management but that its
physical memory control are partially or fully exposed to the untrusted OS, not
to mention other system services an enclave program may require (process
management, I/O, etc.). Even only looking at page tables, we are not sure whether the
known paging attack is the only or the most effective way to extract
sensitive information from the enclave process. In the absence of a
comprehensive understanding of possible attack surfaces, it is not clear whether all proposed protection, which tends to be heavyweight and intrusive,
can even raise the bar to side-channel attacks, and whether an adversary can
switch to a different technique or channel, more lightweight and equally
effective, to extract the information these approaches were designed to
protect.

\vspace{3pt}\noindent\textbf{Understanding memory side channels}. As a first step toward a comprehensive understanding of side-channel threats software vendors face, in this paper, we focus on memory-related side channels, which is an important, and arguably the most effective category of side channels that has ever been studied. Of course, the page-fault side channel falls in the scope of our discussion. To deepen the community's understanding of the memory side-channel attack surfaces and guide the design of defense mechanisms, we believe that it is important to make the following three key points: 
First, \textit{page faults are not the only vector that leaks an enclave program's memory access patterns}. Any follow-up attempt to mitigate memory side-channel leaks should take into account the entire attack surfaces. Second, \textit{not every side-channel attack on enclave induces a large number of Asynchronous Enclave eXits (AEXs) as demonstrated in the page-fault attacks}. This is important because the anomalously high AEX interrupt rate has been considered to be a key feature of SGX side-channel attacks, which can be defeated by the protection designed to capture this signal~\cite{shih:tsgx,Chen:2017:dejavu}. Our finding, however, shows that such interrupt-based protection is fragile, as more sophisticated attacks can avoid producing too many interrupts. Third, \textit{it is possible to acquire a fine-grained side-channel observation, at the cache-line level, into the enclave memory.} Hence, defense that places sensitive code and data on the same page~\cite{Shinde:2015:PYF} will not be effective on the new attacks. 

In this paper, we hope to get across these messages through the following research efforts:


\vspace{3pt}\noindent$\bullet$\textit{ Exploration of memory side-channel attack surfaces}. In our research, we surveyed SGX side-channel attack surfaces involving memory management, identifying 8 types of side-channel attack vectors related to address translation caches in CPU hardware (e.g., TLB, paging-structure caches), page tables located in the main memory, and the entire cache and DRAM hierarchy. This study takes into account each step in the address translation and memory operation, and thus to our knowledge presents the most comprehensive analysis on memory side-channel attack surfaces against SGX enclaves.



\vspace{3pt}\noindent$\bullet$\textit{ Reducing side effects of memory side-channel attacks}.
To demonstrate that a large number of AEXs are \textit{not} a necessary condition of memory side-channel attacks, we develop a new memory-based attack, called \textit{sneaky page monitoring} (SPM). SPM attacks work by setting and resetting a page's \textit{accessed} flag in the page table entry (PTE) to monitor when the page is visited.  Unlike the page-fault attacks~\cite{Xu:2015:controlled}, in which a page fault is generated each time when a page is accessed, manipulation of the \accessed does not trigger any interrupt directly.  However, the attack still needs to flush the \textit{translation lookaside buffer} (TLB) from time to time by triggering interrupts to force the CPU to look up page tables and set \accessed in the PTEs.
Nevertheless, we found that there are several ways to reduce
the number of the interrupts or even completely eliminate them in such attacks. 
Particularly, we can avoid tracking the access patterns between the entry and
exit pages of a program fragment (e.g., a function) and instead, use the
\textit{execution time} measured between these pages to infer the execution
paths in-between.  This approach can still work if all secret-dependent
branches are located in the same page (i.e., a mechanism proposed
by~\cite{Shinde:2015:PYF}), as long as there is still a timing difference
between the executions of these branches. 
Further, we present a technique that utilizes Intel's HyperThreading capability to flush the
TLBs through an attack process sharing the same CPU core with the enclave code, which can eliminate the need of interrupts, when HyperThreading is on.  

We demonstrate the effectiveness of SPM through attacks on real-world applications. Particularly, we show that when attacking EdDSA (Section~\ref{subsec:eddsa}), our timing enhancement only triggers 1,300 interrupts, compared with 71,000 caused by the page-fault attack and 33,000 by the direct \accessed attack, when recovering the whole 512-bit secret key. This level of the interrupt rate makes our attack almost invisible to all known interrupt-based defense~\cite{shih:tsgx,Chen:2017:dejavu}, given the fact that even normal execution of the target program generates thousands of interrupts.



%

\vspace{3pt}\noindent$\bullet$\textit{ Improving attack's spatial granularity}. Page-fault attacks allow attackers to observe the enclave program's memory access pattern at the page granularity (4KB), therefore existing solutions propose to defeat the attacks by aligning sensitive code and data within the same memory pages. To show that this defense strategy is ineffective, we demonstrate a series of memory side-channel attacks that achieve finer spatial granularity, which includes a cross-enclave \primeprobe attack, a cross-enclave DRAMA attack, and a novel cache-DRAM attack. Particularly, the cache-DRAM attack leverages both \primeprobe cache attacks and DRAMA attacks to improve the spatial granularity. When exploiting both channels, we are able to achieve a fine-grained observation (64B as apposed to 16KB for \primeprobe and $\ge$1KB for the DRAMA attack alone), which enables us to monitor the execution flows of an enclave program (similar to \flushreload attacks). Note that this cannot be done at the cache level since in our case the attacker does not share code with the target enclave program, which makes \flushreload impossible.

\vspace{3pt}\noindent\textbf{Implications}. Our findings point to the disturbing
lack of understanding about potential attack surfaces in SGX, which can have a
serious consequence. Not only are all existing defense mechanisms vulnerable to
the new attacks we developed, but some of them only marginally increase the cost
on the attacker side: as demonstrated in our study, for the channels on the
virtual memory, the page-fault attack is not the most cost-effective one and a large number of AEX interrupts are \textit{not} necessary for a successful attack; all existing protection does not
add much burden to a more sophisticated attacker, who can effectively reduce the frequency of AEXs without undermining the effectiveness of the attack. Most importantly, we hope that our study can lead to rethinking
the security limitations of SGX and similar TEE technologies, provoking a
discussion on\ignore{ what they can protect and what they cannot,} when to use them and
how to use them properly.

\vspace{3pt}\noindent\textbf{Contributions}. In this paper, we make the
following contributions:


\vspace{2pt}\noindent$\bullet$\textit{ The first in-depth study on SGX memory side-channel attack surfaces}. Although only focusing on the memory management, our study reveals
new channels that disclose information of the enclave, particularly \accessed, timing and cross-enclave channels.

\vspace{2pt}\noindent$\bullet$\textit{ New attacks}. We developed a suite of new attack techniques that exploit these new channels. Particularly, we show multiple channels can complement each other to enhance the effectiveness of an attack: timing+\accessed to reduce AEXs (rendering existing protection less effective) and DRAM+Cache to achieve a fine-grained observation into the enclave (64B). 

\vspace{2pt}\noindent$\bullet$\textit{ New understanding}. We discuss possible mitigations of the new threats
and highlight the importance of a better understanding of the limitations
of SGX-like technologies.

%% file: tex/background.tex
\section{Background}
\label{sec:background}

\subsection{Memory Isolation in Intel SGX}
\label{subsec:memory}

Memory isolation of enclave programs is a key design feature of Intel SGX. To
maintain backward-compatibility, Intel implements such isolation via extensions
to existing processor architectures, which we introduce below.

\vspace{3pt}\noindent\textbf{Virtual and physical memory management}.  Intel SGX
reserves a range of continuous physical memory exclusively for enclave programs
and their control structures, dubbed Processor Reserved Memory (PRM).  The
extended memory management units (MMU) of the CPU prevents accesses to the PRM
from all programs outside enclaves, including the OS kernel,
virtual machine hypervisors, SMM code or Direct Memory Accesses (DMA).  Enclave
Page Cache (EPC) is a subset of the PRM memory range. The EPC is divided to pages of
4KBs and managed similarly as the rest of the physical memory pages.  Each EPC
page can be allocated to one enclave at one time.

The virtual memory space of each program has an Enclave Linear Address Range
(ELRANGE), which is reserved for enclaves and mapped to the EPC pages. Sensitive
code and data is stored within the ELRANGE.  Page tables responsible for
translating virtual addresses to physical addresses are managed by the untrusted
system software. The translation lookaside buffer (TLB) works for EPC pages in
traditional ways.  When the CPU transitions between \textit{non-enclave mode}
and \textit{enclave mode}, through \texttt{EENTER} or \texttt{EEXIT}
instructions or Asynchronous Enclave Exits (AEXs), TLB entries associated with
the current Process-Context Identifier (PCID) as well as the global identifier
are flushed, preventing non-enclave code learning information about address
translation inside the enclaves.

\vspace{3pt}\noindent\textbf{Security check for memory isolation}.
To prevent system software from arbitrarily controlling the address translation by
manipulating the page table entries, the CPU also consults the Enclave Page Cache Map
(EPCM) during the address translation. Each EPC page corresponds to an entry in
the EPCM, which records the owner enclave of the EPC page, the type of the page,
and a valid bit indicating whether the page has been allocated. When an EPC page
is allocated, its access permissions are specified in its EPCM entry as
\textit{readable}, \textit{writable}, and/or \textit{executable}. The virtual
address (within ELRANGE) mapped to the EPC page is also recorded in the
EPCM entry.

Correctness of page table entries set up by the untrusted system software is
guaranteed by an extended Page Miss Handler (PMH). When the code is executing in
the enclave mode or the address translation result falls into the PRM range,
additional security check will take place. Specially, when the code is running
in the non-enclave mode and address translation falls into the PRM range, or the
code runs in the enclave mode but the physical address is not pointing to a
\textit{regular} EPC page belonging to the current enclave, or the virtual
address triggering the page table walk doesn't match the virtual address
recorded in the corresponding entry in the EPCM, a page fault will occur.
Otherwise, the generated TLB entries will be set according to both the
attributes in the EPCM entry and the page table entry.

\vspace{3pt}\noindent\textbf{Memory encryption}.
To support larger ELRANGE than EPC, EPC pages can be ``swapped'' out to regular
physical memory. This procedure is called EPC page eviction. The confidentiality
and integrity of the evicted pages are guaranteed through authenticated
encryption. The hardware Memory Encryption Engine (MEE) is integrated with the
memory controller and seamlessly encrypts the content of the EPC page that is
evicted to a regular physical memory page. A Message Authentication Code (MAC)
protects the integrity of the encryption and a nonce associated with the evicted
page. The encrypted page can be stored in the main memory or swapped out to
secondary storage similar to regular pages. But the metadata associated with the
encryption needs to be kept by the system software properly for the page to be
``swapped'' into the EPC again.

\subsection{Adversary Model}
In this paper, we consider attacks against enclave-protected code and data. The
system software here refers to the program that operates with system privileges,
such as operating systems and hypervisors. Our focus in this paper is
side-channel analysis that threatens the confidentiality of the enclave
programs. As such, software bugs in the code of an enclave program are out of
our scope. Moreover, side channels not involving memory management and address
translation are not covered either.

We assume in our demonstrated attacks knowledge of the victim binary code to be
loaded into the enclave. As the adversary also knows the base address of the
enclave binary in the virtual address space, as well as the entire
virtual-to-physical mapping, the mapping of the binary code in pages, caches,
DRAMs can be derived. Source code of the victim program is NOT required.  We
conducted analysis and experiments on \textit{real} SGX platforms.  So we do
assume the adversary has access to a machine of the same configuration before
performing the attacks.

%% file: tex/surface.tex
\section{Understanding Attack Surfaces}
\label{sec:surfaces}

In this section, we explore side-channel attack surfaces in SGX memory management, through an in-depth study
of attack vectors (shared resources that allow
interference of the execution inside enclaves), followed by an analysis of
individual vectors, in terms of the way they can be exploited and effectiveness
of the attacks.

\subsection{Attack Vectors}
\label{subsec:vector}

A memory reference in the modern Intel CPU architectures involves a sequence of
micro-operations: the virtual address generated by the program is translated
into the physical address, by first consulting a set of address translation
caches (e.g., TLBs and various paging-structure caches) and then walking through
the page tables in the memory. The resulting physical address is then used to
access the cache (e.g., L1, L2, L3) and DRAM to complete the memory reference.
Here, we discuss memory side-channel attack vectors in each of these steps.

\vspace{3pt}\noindent\textbf{Address Translation Caches.} 
Address translation
caches are hardware caches that facilitate address translation, including TLBs
and various paging-structure caches. TLB is a multi-level set-associative
hardware cache that temporarily stores the translation from virtual page numbers
to physical page numbers.  Specially, the virtual address is first
divided into three components: TLB tag bits, TLB-index bits, and page-offset
bits. The TLB-index bits are used to index a set-associative TLB and the TLB-tag
bits are used to match the tags in each of the TLB entries of the searched TLB
set.  Similar to L1 caches, the L1 TLB for data and instructions are split into
dTLB and iTLB. An L2 TLB, typically larger and unified, will be searched upon L1
TLB misses.  Recent Intel processors allow selectively flushing TLB entries at
context switch. This is enabled by the Process-Context Identifier (PCID) field
in the TLB entries to avoid flushing the entries that will be used again.  If both
levels of TLBs miss, a page table walk will be initiated.  The virtual page
number is divided into, according to Intel's terminology, PML4 bits, PDPTE bits,
PDE bits, and PTE bits, each of which is responsible for indexing one level of
page tables in the main memory.  Due to the long latency of page-table walks, if
the processor is also equipped with paging structure caches, such as PML4 cache,
PDPTE cache, PDE cache, these hardware caches will also be searched to
accelerate the page-table walk.  The PTEs can be first searched in the cache
hierarchy before the memory access~\cite{Glew:1997:MAP}.

\vspace{-0.5em}
\begin{attack}
\label{vector:tlb1}
Shared TLBs and paging-structure caches under HyperThreading.
\end{attack}
\vspace{-0.5em}

When HyperThreading (HT)\footnote{Intel's term for Simultaneous Multi-Threading.} is
enabled, code running in the enclave mode may share the same set of TLBs and
paging-structure caches with code running in non-enclave mode. Therefore, the
enclave code's use of such resources will interfere with that of the non-enclave
code, creating side channels.
This attack vector is utilized to clear the TLB entries in the HT-SPM attack (Section~\ref{sec:htspm}). 

\vspace{-0.5em}
\begin{attack}
\label{vector:tlb2}
Flushing selected entries in TLB and paging-structure caches at AEX.
\end{attack}
\vspace{-0.5em}

According to recent versions of Intel Software Developer's
Manual~\cite{IntelDevelopmentManual}, entering and leaving the enclave mode will
flush entries in TLB and paging-structure caches that are associated with the
current PCID. As such, it enables an adversary from a different process context to infer
the flushed entries at context switch. This is possible even on processors
without HT. However, we were not able to confirm this attack vector on the
machines we had (i.e., Skylake i7-6700). We conjecture that this is because our
Skylake i7-6700 follows the specification in an older version of Intel Software Developer's
Manual~\cite{IntelDevelopmentManualOld}, which states all entries will be
flushed regardless of the process context. Nevertheless, we believe this attack
vector could be present in future processors\ignore{ given the claims in the new manual}.

\vspace{-0.5em}
\begin{attack}
\label{vector:pte-cache}
Referenced PTEs are cached as data.
\end{attack}
\vspace{-0.5em}

Beside paging-structure caches, referenced PTEs will also be cached as regular
data~\cite{Glew:1997:MAP}. This artifact enables a new attack vector: by
exploiting the \flushreload side channel on the monitored PTEs, the adversary can
perform a cross-core attack to trace the page-level memory access pattern of the
enclave code. This attack vector presents a timing-channel version of the sneaky
page monitoring attack we describe in Section~\ref{sec:spm}. We will
discuss its implication\ignore{ to enclave side-channel security} in
Section~\ref{sec:discussion}.

\begin{figure}
\centering
\includegraphics[width=.95\columnwidth]{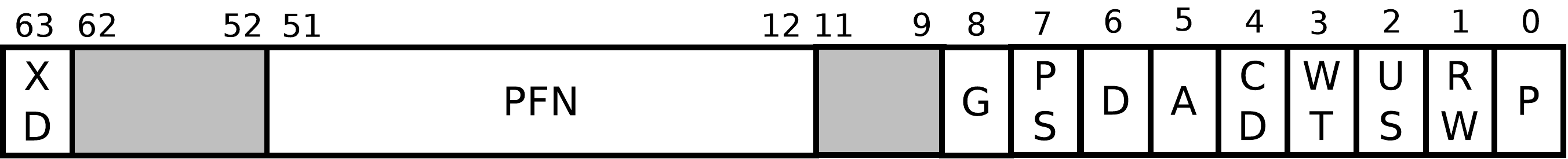}
\caption{Page table entries.}
\label{fig:pte}
\end{figure}

\vspace{3pt}\noindent\textbf{Page tables.}
Page tables are multi-level data
structures stored in main memory, serving address translation. Every page-table
walk involves multiple memory accesses. Different from regular memory accesses,
page-table lookups are triggered by the micro-code of the processor direction,
without involving the re-ordering buffer~\cite{Glew:1997:MAP}. The entry of each
level stores the pointer to (i.e., physical address of) the memory page that
contains the next level of the page table. The structure of a PTE is shown in
Figure~\ref{fig:pte}. Specially, bit 0 is \textit{present} flag, indicating
whether a physical page has been mapped to the virtual page; bit 5 is \textit{accessed} flag,
which is set by the processor every time the page-table walk leads to the
reference of this page table entry; bit 6 is \textit{dirty} flag, which is set
when the corresponding page has been updated. Page frame reclaiming algorithms
rely on the \textit{dirty} flag to make frame reclamation decisions.

As the page tables are located inside the OS kernel and controlled by the
untrusted system software, they can be manipulated to attack enclaves. However,
as mentioned earlier, because the EPC page permission is
also protected by EPCM, malicious system software cannot arbitrarily manipulate
the EPC pages to compromise its integrity and confidentiality. However, it has
been shown in previous work~\cite{Xu:2015:controlled} that by clearing the
\textit{present} flag in the corresponding PTEs, the malicious system software
can collect traces of page accesses from the enclave programs, inferring
secret-dependent control flows or data flows. Nevertheless, setting
\textit{present} flag is not the only attack vector against enclave programs.

\vspace{-0.5em}
\begin{attack}
\label{vector:pte-accessed}
Updates of the \accessed in enclave mode.
\end{attack}
\vspace{-0.5em}

When the page-table walk results in a reference of the PTE, the \textit{accessed} flag of the
entry will be set to 1. As such, code run in non-enclave mode will be able to
detect the page table updates and learn that the corresponding EPC page has
just been visited by the enclave code. However, page-table walk will also
update TLB entries, so that future references to the same page will not update
the \accessed in PTEs, until the TLB entries are evicted by other address
translation activities. We exploit this attack vector in our SPM attacks in
Section~\ref{sec:spm}.

\vspace{-0.5em}
\begin{attack}
\label{vector:pte-dirty}
Updates of the \textit{dirty} flags in enclave mode.
\end{attack}
\vspace{-0.5em}

Similar to \accessed, the \textit{dirty} flag will be updated when the
corresponding EPC page is modified by the enclave program. This artifact can be
exploited to detect memory writes to a new page. The new side-channel attack
vector will enable the adversary to monitor secret-dependent memory writes,
potentially a finer-grained inference attack than memory access tracking.

\vspace{-0.5em}
\begin{attack}
\label{vector:pte-present}
Page faults triggered in enclave mode.
\end{attack}
\vspace{-0.5em}

In addition to the \textit{present} flag, a few other bits in the PTEs can be
exploited to trigger page faults. For example, on a x86-64 processor, bit $M$ to
bit 51 are reserved bits which when set will trigger page fault upon address
translation. Here bit $M-1$ is the highest bit of the physical address on the
machine. The \textit{NX} flag, when set, will force page faults when
instructions are fetched from the corresponding EPC page.

\vspace{3pt}\noindent\textbf{Cache and memory hierarchy.} 
Once the virtual
address is translated into the physical address, the memory reference will be
served from the cache and memory hierarchy. Both are temporary storage that only
hold data when the power is on. On the top of the hierarchy is the separate L1
data and instruction caches, the next level is the unified L2 cache dedicated to
one CPU core, then L3 cache shared by all cores of the CPU package, then the
main memory. Caches are typically built on Static Random-Access Memory (SRAM)
and the main memory on Dynamic Random-Access Memory (DRAM). The upper level
storage tends to be smaller, faster and more expensive, while the lower level
storage is usually larger, slower and a lot cheaper.  Memory fetch goes through
each level from top to bottom; misses in the upper level will lead to accesses
to the next level. Data or code fetched from lower levels usually update entries
in the upper level in order to speed up future references.

The main memory is generally organized in multiple memory channels, handled by memory controllers. One memory channel is
physically partitioned into multiple DIMMs (Dual In-line Memory Module), each
with one or two ranks.  Each rank has several DRAM chips (e.g., 8 or 16), and is
also partitioned into multiple banks. A bank carries the memory arrays organized
in rows and each of the rows typically has a size of 8KB, shared by
multiple 4KB memory pages since one page tends to span over multiple rows. Also
on the bank is a \textit{row buffer} that keeps the most recently accessed row.
Every memory read will load the entire row into the row buffer before the memory
request is served. As such, accesses to the DRAM row already in the row buffer
are much faster.

\vspace{-0.2em}
\begin{attack}
\label{vector:cache}
CPU caches are shared between code in enclave and non-enclave mode.
\end{attack}
\vspace{-0.5em}

SGX does not protect enclave against cache side-channel attacks. Therefore, all levels of caches
are shared between code in enclave mode and non-enclave mode, similar to
cross-process and cross-VM cache sharing that are well-known side-channel attack
vectors. Therefore, all known cache side-channel attacks, including those on L1 data
cache, L1 instruction cache, and L3 cache, all apply to the
enclave settings. We empirically confirmed such threats (Section~\ref{subsec:spatial}).

\vspace{-0.2em}
\begin{attack}
\label{vector:DRAM}
The entire memory hierarchy, including memory controllers, channels, DIMMs, DRAM
ranks and banks (including row buffers), are shared between code in enclave and
non-enclave mode.
\end{attack}
\vspace{-0.5em}

Similar to cache sharing, DRAM modules are shared by all processes running in
the computer systems. Therefore, it is unavoidable to have enclave code and
non-enclave code accessing memory stored in the same DRAM bank.  The DRAM row
buffer can be served as a side-channel attack vector: when the target program
makes a memory reference, the corresponding DRAM row will be loaded into the row
buffer of the bank; the adversary can compare the row-access time to detect
whether a specific row has just been visited, so as to infer the target's memory access.
This artifact has been exploited in DRAMA
attacks~\cite{pessl2016drama}. In Section~\ref{sec:dram}, we show that after key technical challenges are addressed,
such attacks can also succeed on enclave programs. Other shared memory hierarchy can also create
contention between enclave and non-enclave code, causing interference that may
lead to covert channels~\cite{Lampson:1973:NCP}. 

\subsection{Characterizing Memory Vectors}
\label{subsec:model}

Here we characterize the aforementioned memory side-channels in three dimensions:

\vspace{3pt}\noindent\textbf{Spatial granularity}. This concept describes the
smallest unit of information \textit{directly} observable to the adversary
during a memory side-channel attack. Specifically, it measures the size of the
address space one side-channel observation could not reveal.  For example, the
spatial granularity of the page-fault attack is 4KB, indicating that every
fault enables the adversary to see one memory page (4096 bytes) being
touched, though the exact address visited is not directly disclosed.


\vspace{3pt}\noindent\textbf{Temporal observability}. Given a spatial granularity
level, even though the adversary cannot directly see what happens inside the
smallest information unit, still there can be \textit{timing signals} generated
during the execution of the target program to help distinguish different
accesses made by the program within the unit. For example, the duration for a
program to stay on a page indicates, indirectly, whether a single memory or
multiple accesses occur. A side-channel is said to have this property if the
timing is measured and used to refine the observations in the attack.


\vspace{3pt}\noindent\textbf{Side effects}.
We use this concept to characterize observable anomalies caused by a memory
side-channel attack, which could be employed to detect the attack. An example is
AEX, which is frequently invoked by the page-fault attack. Another side effect
is the slowdown of the execution. Since the primary approach to conducting a
side-channel attack is to cause contention in memory resource, such as the flush
of caches, TLBs, paging structure caches, DRAM row buffers, etc., overheads will
be introduced to the runtime performance of the enclave code. AEXs also
contribute to the performance overhead. For example, the original page-fault
attacks are reported to make the target program run one or two orders of
magnitude slower. This level of slowdown is easy to get noticed.  Frequent AEXs
are also detectable using approaches proposed by Chen
\etal~\cite{Chen:2017:dejavu}, as the execution time between two basic blocks
can be much longer.

%% file: tex/attack.tex


\input{tex/attack1}

\input{tex/attack2}

%% file: tex/attack1.tex
\section{Reducing Side Effects with Sneaky Page Monitoring Attacks}
\label{sec:spm}

To attack the virtual memory, a page-fault side-channel attacker first restricts
access to all pages, which induces page faults whenever the enclave process
touches any of these pages, thereby generating a sequence of its page visits. A
problem here is that this approach is heavyweight, causing an interrupt for each
page access.  This often leads to a performance slowdown by one or two orders of
magnitude~\cite{Xu:2015:controlled}.  As a result, such an attack could be
detected by looking at its extremely high frequency of page faults (i.e., AEXs) and
anomalously low performance observed from the remote.  All existing solutions,
except those requiring hardware changes, are either leveraging interrupts or
trying to remove the page trace of a program (e.g., putting all the code on one
page). Little has been done to question whether such defense is sufficient.

To show that excessive AEXs are not the necessary condition to conducting memory side-channel attack, in this section, we elaborate sneaky page monitoring (SPM), a new paging attack that can achieve comparable effectiveness with much less frequent AEXs.   


\subsection{The Three SPM Attacks (Vector 4)}
In this section, we introduce three types of SPM attacks, which monitor the page table entries and exploit different techniques to flush TLBs.

\vspace{3pt}\noindent\textbf{B-SPM: Accessed Flags Monitoring.}
The SPM attack manipulates and
monitors the \accessed on the pages of an enclave process to identify its
execution trace. Specifically, we run a system-level attack process outside an
enclave to repeatedly inspect each page table entry's \textit{accessed} flag, record when it
is set (from 0 to 1) and reset it once this happens. The page-access trace
recovered in this way is a sequence of \textit{page sets}, each of which is a
group of pages visited (with their \accessed set to 1) between two consecutive
inspections. This attack is lightweight since it does not need any AEX to observe the pages first time when they are visited.

\begin{figure}
\centering
\includegraphics[width=.9\columnwidth]{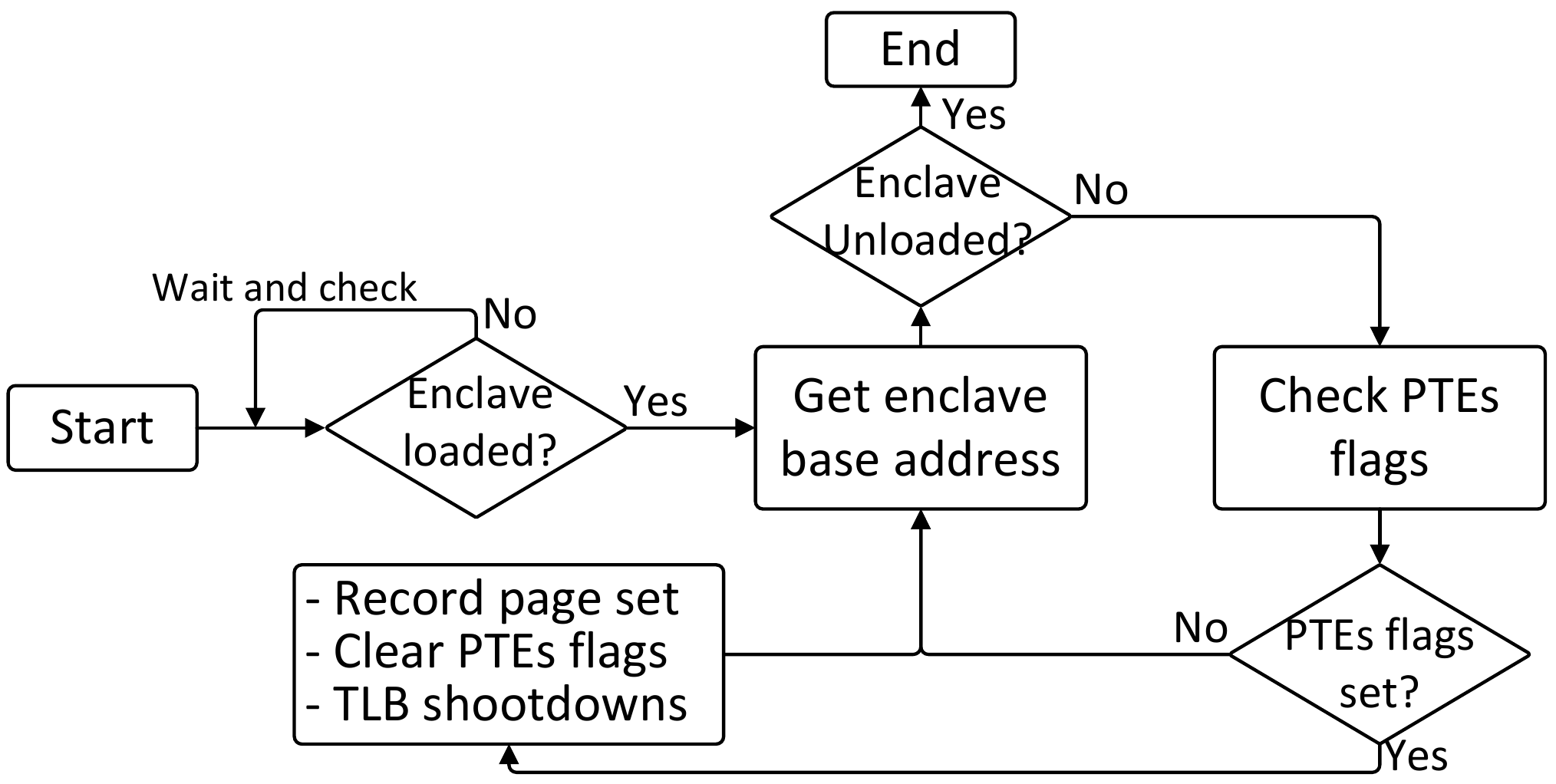}
\caption{Basic SPM attack.}
\label{fig:basic}
\end{figure}

However, as mentioned earlier, after a virtual address is translated, its page number is automatically added to a TLB. As a result, the \textit{accessed} flag of that
page will not be set again when the page is visited later. To force the processor to
access the PTE (and update the flag), the attacker has to invalidate the TLB
entry proactively.  The simplest way to do so is by generating an
inter-processor interrupt (IPI) from a different CPU core to trigger a TLB
shootdown, which causes an AEX from the enclave, resulting in flushing of all
TLB entries of the current PCID. Figure~\ref{fig:basic} illustrates this attack,
which we call \textit{basic SPM} or \textit{B-SPM}.

This B-SPM attack still involves interrupts but is already much more
lightweight than the page-fault attack: TLB shootdowns are typically less expensive
than page faults; more importantly, B-SPM only triggers interrupts \textit{when
visiting the same page needs to be observed again}, while the latter needs to
trigger an interrupt for \textit{every (new) page access}. 


In terms of accuracy, the page-fault attack tends to have a finer-grained
observation while B-SPM attack cannot differentiate the visiting order of two pages that
are spotted during the same round of inspections. However B-SPM attack strives for a balance between the interrupt rate and attack resolutions.


\ignore{In our experiment, we implemented kernel thread within a Linux module which monitored the loading of victim enclave every 15 milliseconds.}

\vspace{3pt}\noindent\textbf{T-SPM: Timing enhancement.} 
\begin{figure}
\centering
\includegraphics[width=.6\columnwidth]{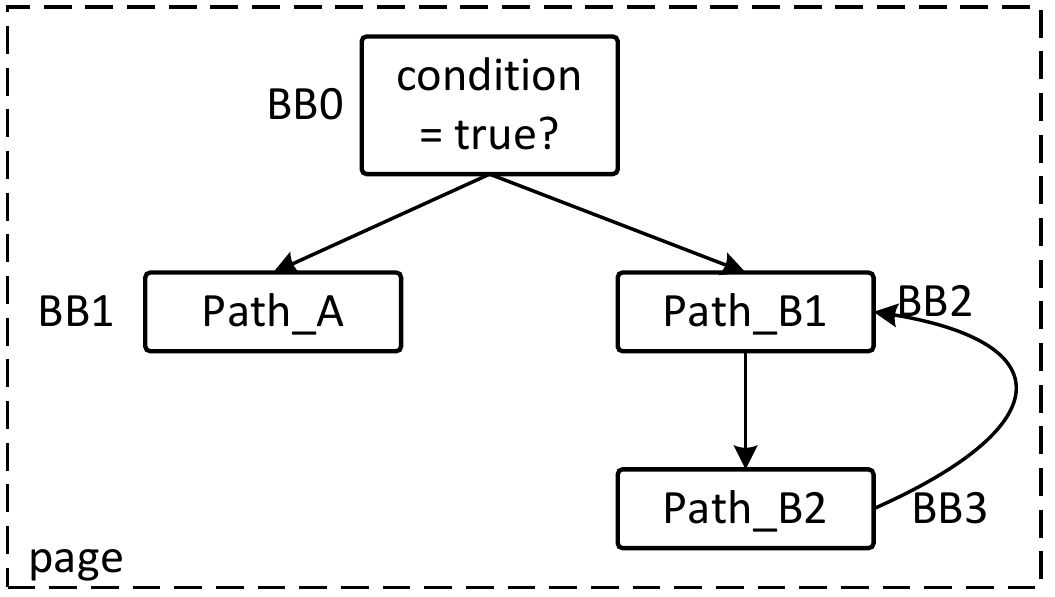}
\caption{An example of secret-dependent branch leaking timing information.}\label{lst:timing}
\end{figure}
When repeated visits to same pages become a salient feature for an input, the basic SPM needs to issue more TLB shootdowns in order to observe the feature, making the attack observable to the existing protections that detect the anomalous interrupt rate~\cite{shih:tsgx,Chen:2017:dejavu}. Figure~\ref{lst:timing} illustrates an example, in which the secret-dependent code resides in the same page, except that the execution on one condition involves a loop while that on the other does not, leading to different execution time. In this case, TLB shootdowns during the execution of the loop are required to distinguish two branches using page visit traces (i.e. number of page visits). To reduce the number of the interrupts, we leverage a timing channel to enhance
SPM, making it stealthier. Specifically, given a code fragment with a unique
entry page $\alpha$ and a unique exit page $\beta$, together with multiple
input-dependent paths between the two points on different pages, our
timing-enhanced SPM (called \textit{T-SPM}) continuously monitors $\alpha$ and
$\beta$, measuring the execution time between these two points, and once the
\textit{accessed} flag of $\beta$ is found to be set, flushes the TLB and resets the
\accessed for both PTEs. The timing recorded is then used to infer the input of
the code fragment.

This simple approach avoids all the interrupts between $\alpha$ and $\beta$, but
still reveals the possible execution path connecting them. In the extreme case,
when all other code stays on the same page, as proposed by the prior
research~\cite{Shinde:2015:PYF} to defend against page-fault attacks, T-SPM can
still infer sensitive information when the operations on them take different time to
complete.

\vspace{3pt}\noindent\textbf{HT-SPM: TLB Flushing through HyperThreading.}\label{sec:htspm}  Further we
found that when HyperThreading is turned on for a processor, we can clear up the
TLBs without issuing TLB shootdowns, which renders all existing interrupt-based
protection ineffective. HyperThreading runs two virtual cores on a
physical core to handle the workloads from two different OS processes. This
resource-sharing is transparent to the OS and therefore does not trigger any
interrupt. The processes running on the two virtual cores share some of the TLBs,
which allows the attacker to remove some TLB entries outside the enclave,
without causing any interrupts. As a result, in the presence of
HyperThreading, we can run an attack process together with an enclave
process, to continuously probe the virtual addresses in conflict with the TLB
entries the latter uses, in an attempt to evict these entries and force the
victim process to walk its page tables. Using this technique, which we call
\textit{HT-SPM}, we can remove most or even eliminate the
interrupts during an attack.

\begin{table}
\centering 
\footnotesize{
\caption{Configuration of the testbed, \revise{available per logical core when HyperThreading is enabled}.}
\label{tab:configuration}
\begin{tabular}[t]{c|c|c}
\Xhline{1pt}
& \textbf{Size} & \textbf{Sets $\times$ Ways} \\
\Xhline{0.5pt}
\textbf{iTLB} & 64 & 8 $\times$ 8 \\
\textbf{dTLB} & 64 & 16 $\times$ 4\\
\textbf{L2 TLB} & 1536 & 128 $\times$ 12\\
\textbf{iCache} & 32KB & 64 $\times$ 8\\
\textbf{dCache} & 32KB & 64 $\times$ 8\\
\textbf{L2 Cache} & 256KB &  1024 $\times$ 4\\
\textbf{L3 Cache} & 8MB & 8192 $\times$ 16\\
\Xhline{1pt}
& \textbf{Size} & \textbf{Channel $\times$ DIMMs $\times$ Ranks $\times$ Banks
$\times$ Rows}\\
\Xhline{0.5pt}
\textbf{DRAM} & 8GB$\times$2 & $2\times 1 \times 2 \times 16 \times 2^{15}$ \\
\Xhline{1pt}
\end{tabular}} 

\end{table}

\subsection{Evaluation of Effectiveness}

Our analysis was performed on an Dell Optiplex 7040 with a Skylake
i7-6700 processor and 4 physical cores, with 16GB memory. The configuration of
the cache and memory hierarchy is shown in Table~\ref{tab:configuration}. It runs
Ubuntu 16.04 with kernel version 4.2.8. During our experiments, we patched the
OS when necessary to facilitate the attacks, as an OS-level adversary would do. We used the latest Graphene-SGX Library OS~\cite{github.graphene.sgx,tsai2014cooperation} compiled using GCC 5.4.0 with default compiling options to port unmodified libraries.

\vspace{3pt}\noindent\textbf{B-SPM on Hunspell.}
Hunspell is a popular spell checking tool used by software packages like Apple's OS X and
Google Chrome. It stores a dictionary in a hash table, which uses linked lists
to link the words with the same hash values. Each linked list spans across
multiple pages, so searching for a word often generates a unique page-visit
sequence. Prior study~\cite{Xu:2015:controlled} shows that by monitoring page
faults, the attacker outside an enclave can fingerprint the dictionary lookup
function inside the enclave, and further determine the word being checked from
the sequence of accessing different data pages (for storing the dictionary).  In
our research, we evaluated B-SPM on Hunspell 1.3.3 and found that the invocation
of its \texttt{spell} function (looking up for one word) can be determined by
the access of a specific page, which can be reliably identified at the
inspection rate (for an attack process running on a separate core) of once per
184 CPU cycles.  For simplicity, our attack process issues a TLB shootdown once
the function invocation is discovered. In the interrupt, the process inspects
the PTEs of pages being monitored to identify the searched word and resets their
\accessed, and then monitors the occurrence of the next function invocation.
This approach identifies all the iterative lookups for multiple
words.

\begin{table}
\centering
{\footnotesize
\caption{Words distribution in the en\_US Hunspell dictionary.}\label{table:distribution}
\begin{tabular}{c|c|c|c|c}
\Xhline{1pt}
\multirow{2}{*}{\textbf{group size}} &
\multicolumn{2}{c|}{\textbf{Page-fault based}} &
\multicolumn{2}{c}{\textbf{Accessed-flag based}}   \\
\cline{2-5}
& words & \% & words &\%
\\ \Xhline{0.5pt}
\textbf{1} & 51599 & 83.05 & 45649 & 73.47 \\
\textbf{2} & 7586 & 12.21 & 8524 & 13.72 \\
\textbf{3} & 2073 & 3.34 & 3027 & 4.87 \\
\textbf{4} & 568 & 0.91 & 1596 & 2.57 \\
\textbf{5} & 200 & 0.32 & 980 & 1.58 \\
\textbf{6} & 60 & 0.10 & 810 & 1.30 \\
\textbf{7} & 35 & 0.06 & 476 & 0.77 \\
\textbf{8} & 8 & 0.01 & 448 & 0.72 \\
\textbf{9} & 0 & 0 & 306 & 0.49 \\
\textbf{10} & 0 & 0 & 140 & 0.23 \\
\textbf{> 10} & 0 & 0 & 173 & 0.28 \\
\Xhline{1pt}
\end{tabular}
}
\end{table}

Like the prior research~\cite{Xu:2015:controlled}, we also evaluated our attack
using the \path{en_US} Hunspell dictionary, as illustrated in
Table~\ref{table:distribution}. To compare with the page-fault attack, we
re-implemented it and ran it against the same data-set, whose results are shown
in Table~\ref{table:distribution}. As we can see here, the effectiveness of
B-SPM is in line with that of the known attack: e.g., the percentage of
the uniquely-identifiable words (i.e., group size 1) is 73.47\% in our attack, a
little below 83.05\% observed in the page-fault attack; more than 92\% of the
words are in group size less than or equal to 3, compared with 98.6\% in the
page-fault attack. When it comes to performance, however, B-SPM runs much
faster: for 62,129 word look-ups it slowed down the original program by a factor
of 5.1$\times$, while the existing attack incurred an overhead of
1214.9$\times$. Note that the prior research reports a slowdown of 25.1$\times$
for 39,719 word look-ups over the SGX emulator~\cite{Xu:2015:controlled}. In our
study, however, we ran both experiments on the \textit{real} SGX platform.


\vspace{3pt}\noindent\textbf{T-SPM on FreeType.}
FreeType is a font library that converts text content (the input) into images, which has been used by Linux, Android and iOS and other software packages. In our research, we ran T-SPM on its TrueType font rendering function, as did in the prior study~\cite{Xu:2015:controlled}.  The function, \texttt{TT\_load\_Glyph}, takes a letter's glyph as its input to construct its bitmap. The prior study fingerprints the start and the end of the function, and selects a set of pages in-between and uses the number of page faults observed on these pages to determine the letter being rendered.  In our research, we utilize a trigger page to identify the execution of the \path{TT_load_Glyph} function and then within the function, select 5 different $\alpha$-$\beta$ pairs along its control-flow graph as features for identifying the 26 alphabet and the space between words~(see Table~\ref{table:freetypefeature}). Each feature, the timing between its $\alpha$ and $\beta$ points, can separate some of these 27 letters from others.  Collectively, they form a feature vector over which we run a Random Forest Classifier (with number of estimators set as 400) to classify an observed execution of \texttt{TT\_load\_Glyph} into one of these letters.

\begin{table}
\centering
{\footnotesize
\caption{Features used in Freetype experiment.}
\label{table:freetypefeature}
\begin{tabular}{m{0.6in}<{\centering}|m{1.2in}<{\centering}}
\Xhline{1pt}
trigger page & 0x0005B000\\
\hline
\multirow{5}{*}{\textbf{$\alpha$-$\beta$ pairs}} & 0005B000, 0005B000   \\
& 0005B000, 00065000  \\
& 0005B000, 0005E000 \\
& 00065000, 00022000 \\ 
& 0005E000, 00018000 \\
\Xhline{1pt}
\end{tabular}
}
\end{table}

\begin{table}
\centering
{\footnotesize
\caption{T-SPM attack on Freetype 2.5.3: for example, we achieved a precision of 69.90\% over a coverage of 100\% characters.}\label{table:freetype}
\begin{tabular}{c|c|c|c|c|c}
\Xhline{1pt} 
\textbf{coverage} & 100\% & 88.17\% & 75.62\% & 69.14\% & 57.35\% \\ \Xhline{0.5pt}
\textbf{precision} & 69.90\% & 75.25\% & 80.66\% & 84.45\% & 89.94\% \\
\Xhline{1pt}
\end{tabular}
}
\end{table}

We ran our experiment on FreeType 2.5.3 within an enclave and collected 250 samples of a 1000 character paragraph from the book \textit{The Princess and the Goblin} as a training set for the Random Forest Classifier.  Then we tested on a 1000 character paragraph from \textit{The Wonderful Wizard of Oz}, as is used in the prior study~\cite{Xu:2015:controlled}. Based upon the timing vectors observed in the experiments (with an inspection rate of once per 482 cycles), our classifier correctly identified 57.35\% of the characters with a precision of 89.94\% and 100\% of the characters with a precision of 69.90\% (see Table~\ref{table:freetype}). \revise{Particularly, all \textit{space} characters were correctly identified with no false positives.} 72.14\% of the words were correctly recovered by running a dictionary spelling check. Compared with the page-fault attack, which captured 100\% of the words, T-SPM is less accurate but much more efficient: it incurred an overhead of 16\%, while our re-implemented page-fault attack caused the program to slow down by a factor of 252$\times$.

\vspace{3pt}\noindent\textbf{HT-SPM on Hunspell.}
As an example, we ran HT-SPM on Hunspell, in a scenario when a set of words were queried on the dictionary. We conducted the experiments on the Intel Skylake i7-6700 processor, which is characterized by multi-level TLBs (see Table~\ref{tab:configuration}). The experiments show that the dTLB and L2 TLB are fully \textit{shared} across logical cores. Our attack process includes 6 threads:
2 \textit{cleaners} operating on the same physical core as the Hunspell process
in the enclave for evicting its TLB entries and 4 \textit{collectors} for
inspecting the \accessed of memory pages. The cleaners probed all 64 and 1536
entries of the dTLB and L2 TLB every 4978 cycles and the collectors inspected the
PTEs once every 128 cycles. In the experiment, we let Hunspell check 100 words
inside the enclave, with the attack process running outside. The collectors,
once seeing the fingerprint of the \texttt{spell} function, continuously
gathered traces for data-page visits, from which we successfully recovered the
exact page visit set for 88 words.\ignore{The attack is slightly less
effective than B-SPM due to its missing of some page visits under the TLB
clean-up rate and the lack of the boundary information between words (which can
be observed from the instruction pages).} The attack incurred a slowdown of
39.1\% and did not fire a single TLB shootdown.


\subsection{Silent Attacks on EdDSA}
\label{subsec:eddsa}

To understand the stealthiness of different attacks, in terms of their AEX frequency (which are used by the prior research to detect page side-channel attacks~\cite{shih:tsgx,Chen:2017:dejavu}), we ran the page-fault attack, B-SPM and T-SPM against the latest version of Libgcrypt (v1.7.6) to recover the EdDSA session keys\footnote{The attacks only involve code pages, while HT-SPM is designed to reduce AEXs for data pages. As such, HT-SPM is not presented in the comparison.}. Edwards-curve Digital Signature Algorithm (EdDSA)~\cite{bernstein2012high} is a high-speed high-security digital signature scheme over twisted Edwards curves. The security of EdDSA is based on the difficulty of the well-known elliptic curve discrete logarithm problem: given points P and Q on a curve to find an integer $a$, if it exists, such that $Q = aP$. Provided the  security parameters $b$ and a cryptographic hash function $H$ producing $2b$-bit output, an EdDSA secret is a $b$-bit string $k$, and $a=H(k)$ is also private. The corresponding public key is $A = sB$, with $B$ the base point and $s$ the least significant $b$ bits of $a$. Let $r$ be the private session key, the signature of a message $M$ under $k$ is a $2b$-bit string $(R, S)$, where $R = rB$ and $S = (r+H(R, A, M)a)\bmod l$. It can be seen that if $r$ is disclosed, assuming $H(R, A, M)\bmod l \neq 0$, the long-term secret key $a$ can be directly obtained as $a = (S-r)/H(R, A, M)\bmod l$.

\begin{figure}
\centering
\begin{lstlisting}
void
_gcry_mpi_ec_mul_point (mpi_point_t result,
                        gcry_mpi_t scalar, mpi_point_t point,
                        mpi_ec_t ctx) {
	if (ctx->model == MPI_EC_EDWARDS
      || (ctx->model == MPI_EC_WEIERSTRASS
          && mpi_is_secure (scalar))) {
		if (mpi_is_secure (scalar)) {
			/* If SCALAR is in secure memory we assume that it is the secret key we use constant time operation.  */
			...
		}
		else {
			for (j=nbits-1; j >= 0; j--) {
				_gcry_mpi_ec_dup_point (result, result, ctx);
				if (mpi_test_bit (scalar, j))
					_gcry_mpi_ec_add_points (result, result, point, ctx);
			}
		}
		return;      
	}
}
\end{lstlisting}
\caption{Scalar point multiplication for ECC.}\label{lst:eddsa_mul}
\end{figure}

\begin{table}
\centering
{\footnotesize
\caption{Attack summary on EdDSA (Libgcrypt 1.7.5). A normal execution of EdDSA signature without attack also incurs over 1500 AEXs.}\label{table:eddsa}
\begin{tabular}{m{0.8in}<{\centering}|m{1.1in}<{\centering}|c}
\Xhline{1pt}
& \textbf{Monitored pages} & \textbf{Number of AEXs}
\\ \Xhline{0.5pt}
\multirow{2}{*}{\textbf{Page fault attack}} & 000E7000, 000E8000  & \multirow{2}{*}{71,000}\\
& 000F0000, 000F1000 & \\
\hline
\multirow{2}{*}{\textbf{B-SPM attack}} & 000EF000 (trigger page)  & \multirow{2}{*}{33,000}\\
& 000E9000, 000F0000 & \\
\hline
\multirow{2}{*}{\textbf{T-SPM attack}} & 000F0000 (trigger page)  & \multirow{2}{*}{1,300}\\
& 000F1000 (trigger page) & \\
\Xhline{1pt}
\end{tabular}
}
\vspace{-0.06in}
\end{table}

Figure~\ref{lst:eddsa_mul} presents the main function for ECC scalar point multiplication. Although Libgcrypt provides side-channel protection by tagging the long-term private key as ``secure memory'', we found that it does not protect the secret session key. As a result, the non-hardened branch of line 13-17 is always taken while generating the message signature. Then the secret-dependent $if$-branch can leak out session key information. We present our evaluation results using page fault attack, B-SPM and T-SPM respectively, as follows:

\vspace{3pt}\noindent\textbf{Page-fault attacks.} During an offline analysis, we generated the sub-function call traces for both \path{_gcry_mpi_ec_dup_point} (Line 14 of Figure~\ref{lst:eddsa_mul}) and \path{_gcry_mpi_ec_add_points} (Line 16, a necessary condition for bit 1), from which we identified 4 code pages to be monitored, including \path{_gcry_mpi_ec_mul_point} on one page, \path{_gcry_mpi_ec_add_points} and  \path{_gcry_mpi_ec_dup_point} on another page, their related functions on the third page and \path{_gcry_mpi_test_bit} on the last page, whose execution indicates the end of the processing on the current bit. During the attack, we intercepted the page fault handler in our kernel module and restricted accesses to these monitored pages by clearing their present bits. Once a page fault on a monitored page occurred we reset its present bit and recorded the current count of page faults. We found that for key bit $1$ and $0$, there are 89 and 48 subsequent page visits respectively. In total around 71,000 page faults were triggered to correctly recover all the session key bits.

\vspace{3pt}\noindent\textbf{B-SPM attacks.} We found that the aforementioned code page set is visited for both key bit $1$ and $0$, if we do not flush the TLB.  Therefore, the spying thread needs to interrupt the target enclave thread and clean up the current TLB to get more detailed information about page visits to differentiate the key bit with different values. To reduce the frequency of the interrupts needed, instead of sending IPIs with fixed time interval, the spying thread runs simultaneously with the target thread and monitors a trigger page containing \path{ec_pow2} and \path{ec_mul2}. Whenever the trigger page is accessed, the spying thread interrupts the target thread to shoot down the TLB, and then identifies whether two other pages in the page set ($000E9000$ and $000F0000$ in Table~\ref{table:eddsa}) are visited between two interrupts.  We observed a clear difference in the page traces for key bit $1$ and $0$ and can recover all key bits during the post-processing phase. In total around 33,000 interrupts were triggered to correctly recover all the session key bits.

\vspace{3pt}\noindent\textbf{T-SPM attacks.} To further reduce the AEX frequency, we monitor the 2 pages containing \path{_gcry_mpi_ec_mul_point} and \path{_gcry_mpi_ec_dup_point}/\path{_gcry_mpi_ec_add_points} respectively and utilize the time between the visits to both pages to find out the value of the current key bit. Specifically, once both of them are found to be accessed, our attack process starts the timer (using \path{rdtsc}) but waits for 2000 nanoseconds to ensure that the execution of the target process leaves both pages, before shooting down the TLB and resetting the \accessed of both pages. The timer stops when both pages are observed again. In this way, only about 2 interrupts are needed for collecting information for each key bit.  The recorded timings turn out to be differentiating enough to determine whether\path{_gcry_mpi_ec_dup_point} or  \path{_gcry_mpi_ec_add_points} has been executed, around 19,700 cpu cycles for the former and 27,900 cpu cycles for the latter. After all the call traces are gathered, we can figure out that the current key bit is 1 when \path{_gcry_mpi_ec_add_points} is observed right after \path{_gcry_mpi_ec_dup_point}, and 0 if only \path{_gcry_mpi_ec_dup_point} is seen. In total around 1,300 interrupts were triggered to correctly recover all the session key bits.


In summary, we found that these three attacks all are able to recover the full EdDSA session key reliably. Page fault attack triggers a page fault for every page observation and produces about 71,000 AEXs. The B-SPM attack can observe the page visit set between two consecutive inspections. However it still needs to aggressively send IPIs to clear TLB entries to gain timely observation of the pages visited, which produces about 33,000 AEXs. T-SPM attack only issues a TLB shootdown for every invocation of \path{_gcry_mpi_ec_dup_point} or \path{_gcry_mpi_ec_add_points} and differentiates between the two functions using timing information. As such, it generates a minimum number of AEXs. We noticed that a normal execution of the EdDSA program also incurs at least 1,500 AEXs. The OS attacker could reduce the number of additional AEXs (e.g., only 1300 AEXs for T-SPM in the demonstrated example) caused by normal page faults and interrupts, and therefore make the T-SPM attack unobservable.

%% file: tex/attack2.tex
\section{Improving Spatial Granularity with Cache-DRAM Attacks}
\label{sec:dram}

Page-fault side-channel attacks (and also the sneaky page monitoring attacks described in the previous section) only allow attackers to learn the enclave program's memory access patterns at a page granularity. Therefore, mechanisms that mix sensitive code and data into the same pages have been proposed to defeat such attacks~\cite{Shinde:2015:PYF}. Intel also recommends ``aligning specific code and data blocks to exist entirely within a single page.''~\cite{enclaveguide}. However, the effectiveness of this defense method is heavily conditioned on the fact that page granularity is the best spatial granularity achievable by the adversary. However, our study suggests it is not the case.

In this section, we demonstrate three attack methods to show that a powerful adversary is able to improve spatial granularity significantly. Particularly, we will demonstrate a cross-enclave  Prime+Probe cache attack, a cross-enclave DRAMA attack, and a cache-DRAM attack. Because SGX do not allow memory sharing between enclaves, the Flush+Reload cache attacks that can achieve cache-line granularity cannot be conducted against secure enclaves. However, we show that the cache-DRAM attack is capable of achieving the same level of spatial granularity against enclaves.

\subsection{Cross-enclave Prime+Probe (Vector 7)}
\label{subsec:spatial}

Our exploration starts with a validation of cross-enclave cache Prime+Probe attack. SGX is not designed to deal with cache side-channel attacks. Therefore, it is expected that known cache attacks also work against SGX enclaves. 
To confirm this, we ported
GnuPG 1.4.13 to Graphene-SGX. The
algorithm repeatedly decrypted a ciphertext which was encrypted with a 3,072-bit
ElGamal public key, just as the prior work (i.e.,~\cite{Yarom:2015:LLC}) did. GnuPG uses Wiener's table to decide subgroup sizes matching field sizes and adds a a 50\% margin to the security parameters, consequently a private key of 403 bits is used. In the experiment an attack
process monitored when the victim enclave was loaded and determined the physical
address of Square-and-Multiply exponentiation. With knowledge of cache slicing
and cache set mapping~\cite{irazoqui2015systematic}, the attacker constructed
eviction sets mapped to the same cache sets as the target addresses. In
our experiment, with the observation of only one ElGamal decryption, we could
recover all 403 bits of the private key through a \primeprobe cache attack
with an error rate of 2.3\%.

This experiment suggest that Prime+Probe cache attacks can be performed in a cross-enclave scenario, similar to the traditional settings. We note that Prime+Probe attacks achieves a spatial granularity of a cache set, which is 16KB on a processor with a 8196-set LLC (see Table~\ref{tab:configuration} and Table~\ref{tab:analysis}).

\ignore{
In this section, we describe a novel cache-DRAM attack, which achieves the 64 bytes spatial granularity. Worth notable is the fact that this cannot be available in traditional cache timing framework without the assumption of shared memory between the spying thread and the victim thread. It takes advantage of both vector 7 and vector 8, and the ability of the attacker for virtual address to physical address mapping under the SGX threat model.

A successful cache-DRAM attack needs to overcome the following obstacles. First, the victim memory accesses arrives at a DRAM cell; Second, the spying thread knows the DRAM row that the victim will access; Third, detecting a DRAM row access. The first problem can be solved by simply disabling cache behavior on the core running victim thread, however this will incur drastic slow down to the victim thread as well as bring noises to the DRAM row access. A better option is thus adopting a probing thread keeping probing the target cache set. The second problem can be easily solved considering the attacker knows the mapping of virtual to physical addresses, along with the mapping of physical addresses to DRAM cells. The third problem can be solved by observing the timing difference between a DRAM row hit and row miss.}

\subsection{Cross-enclave DRAMA (Vector 8)} 

The DRAMA attack exploits shared DRAM rows to extract
sensitive information~\cite{pessl2016drama}. In such an attack, in order to learn whether the victim process has accessed a virtual address $d$, the adversary allocates two memory blocks that map to the same DRAM bank, with one sharing the same DRAM row with the physical memory of $d$, which we call $p$, and the other mapped to a different row on the same bank, which we call $p'$. The attack is conducted using the following steps:


\vspace{1pt}$\bullet$  Access the memory block $p'$.

\vspace{1pt}$\bullet$  Wait for some victim operations.

\vspace{1pt}$\bullet$  Measure the access time of memory block $p$.


A faster memory access to memory block $p$ suggests the victim process has
probably touched memory address $d$ during its operations. Of course, because
the DRAM row is large (e.g., typically 8 KB), false detection is likely. Even
so, DRAMA is shown to effectively detect the existence of keystroke
activities~\cite{pessl2016drama}.

Directly applying DRAMA to perform cross-enclave attacks faces several challenges, most of which are also faced by our design of cache-DRAM attacks. Therefore, we defer the discussion of these design challenges to Section~\ref{sec:spatial:cachedram} where we detail the cache-DRAM attacks. Here we enumerate some limitations of cross-enclave DRAMA attacks.

{\em First}, most of the victim's memory access will
be cached (EPC is cacheable by default), and hence no information will be leaked
through the use of DRAM rows. 
While we could manually disable cache by setting the cache disable (CD) bit of CR0 for the core running the victim enclave\footnote{In Intel SGX programming reference~\cite{intelprogramref} it is said that
PRMRR\_BASE register could be programmed with values UC(0x0) to set PRM range as uncacheable. We confirmed on our platform that PRMRR\_BASE register cannot be changed after system boot.}, this would slow down the enclave process for approximately $1000\times$.

{\em Second}, DRAMA attacks may falsely detect row hits that are unrelated to the victim enclave's visit to $d$, because the 8KB DRAM row can be shared by multiple data structure or code regions. This false detection, however, is very common in our experiments. 



{\em Finally}, DRAMA cannot achieve fine-grained spatial accuracy. As an example, on our test system a memory page
is distributed over 4 DRAM rows. In an extreme case the attacker could occupy an entire row except a single 1KB chunk for the victim enclave and achieve a spatial accuracy of 1KB (see Table~\ref{tab:analysis}), which is better than the \primeprobe cache attack (16 KB), however still worse than a \flushreload cache attack (64B).

\subsection{Cache-DRAM Attacks (Vector 7 \& 8)}
\label{sec:spatial:cachedram}
To improve cross-enclave DRAMA attacks, we propose a novel cache-DRAM attack. We show that by leveraging both vector 7 and 8, the adversary can significantly improve the spatial granularity of memory side-channel attacks.

\begin{figure}
\centering
\includegraphics[width=.65\columnwidth]{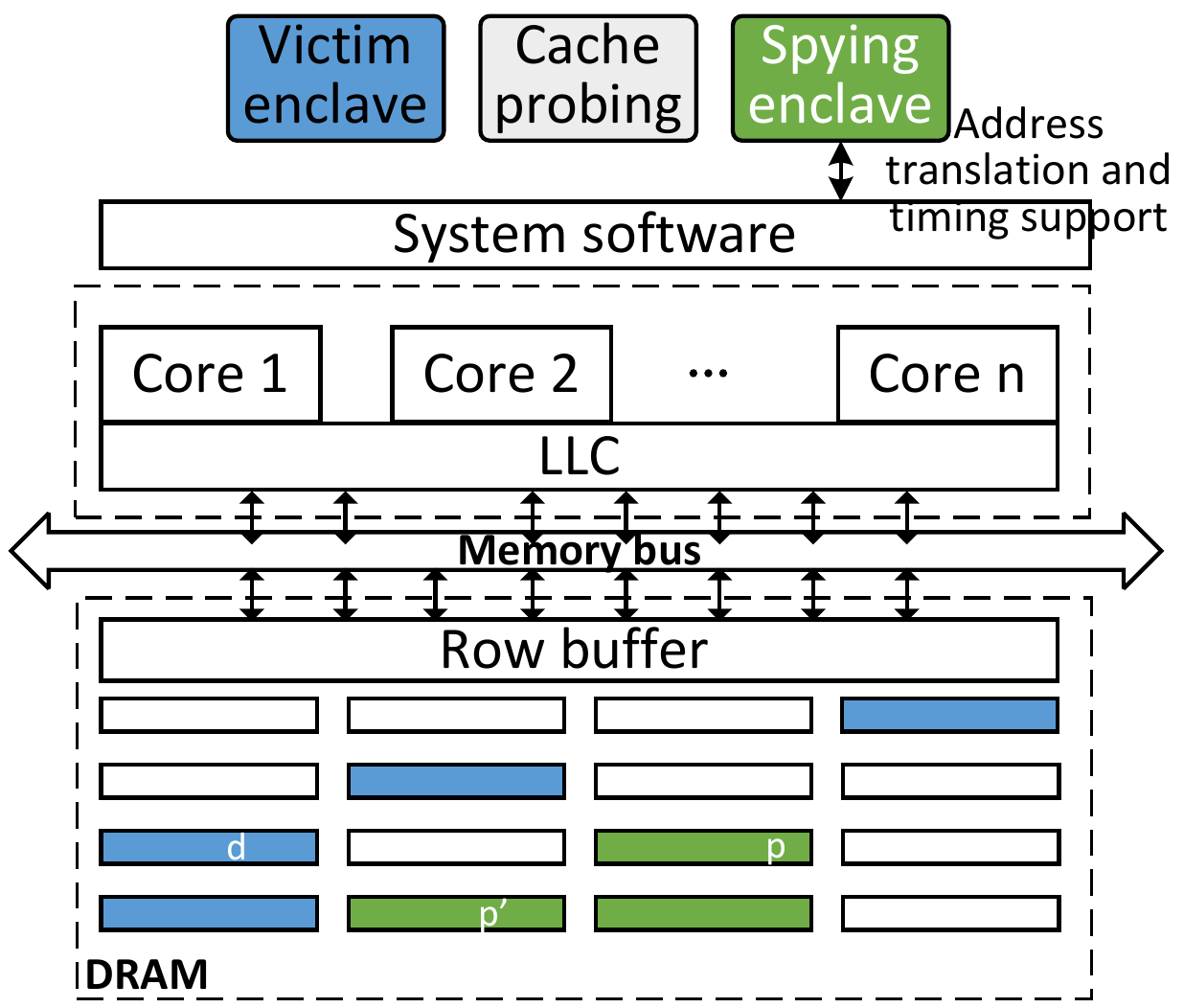}
\caption{Illustration of cache-DRAM attack.}
\label{fig:drama}
\end{figure}

\vspace{3pt}\noindent\textbf{Techniques.}
Particularly, the cache-DRAM attack is performed using two threads: one thread runs in non-enclave mode which \primeprobe{s} a cache set in the last-level cache in which the address $d$ is mapped; the other thread conducts the cross-enclave DRAMA without disabling caching. As the \primeprobe attack causes conflicts with $d$ in the last-level cache, the victim enclave's accesses of $d$ will reach the DRAM. The concept of cache-DRAM attack is shown in Figure~\ref{fig:drama}. However, to implement cache-DRAM attacks against SGX enclaves, one needs to address the following challenges:

\textit{First, share the DRAM Bank and Row with $d$.}
The EPC memory exclusively used by enclaves is already isolated from the rest of the physical memory in DRAMs. To understand this artifact, we explain the mechanism of the DRAM-level isolation using our own testbed (Table~\ref{tab:configuration}) as an example.  With the assumption of row bits being the most significant bits in a physical address~\cite{pessl2016drama, Xiao:2016:OBF}, any bit beyond bit 19 is a row bit that determines the corresponding DRAM row of the physical address.  With a 128MB PRM (physical memory range 0x80000000 to 0x87FFFFFF), no non-PRM memory will occupy row number 0x1000 to 0x10FF, as shown in Table~\ref{table:rowranges}. As such, the PRM range (exclusively taken by enclaves) spans every DRAM bank and occupies specific sets of rows in each bank; these rows are not shared with non-PRM memory regions.

\begin{table}
\centering 
\footnotesize{
\caption{Row ranges for different PRM size.}\label{table:rowranges}
\begin{tabular}[t]{c|c|c}
\Xhline{1pt}
\textbf{PRM size} & \textbf{PRM range} & \textbf{DRAM row range} \\
\Xhline{0.5pt}
32MB & 0x88000000$\sim$0x89FFFFFF & 0x1100$\sim$0x113F \\
64MB & 0x88000000$\sim$0x8BFFFFFF & 0x1100$\sim$0x117F \\
128MB & 0x80000000$\sim$0x87FFFFFF & 0x1000$\sim$0x10FF \\
\Xhline{1pt}
\end{tabular}}
\end{table}


To overcome this barrier, we leverage the processor's support for running multiple enclave programs concurrently to carry out the DRAMA attacks from another enclave program controlled by the adversary. Since both programs operate inside enclaves, they share the EPC memory range. The adversary can manage to co-locate the memory page with the target enclave memory on the same banks and even the same rows, as illustrated in Figure~\ref{fig:drama}.  Specifically, we first identified the physical address of interest in the victim enclave. This can be achieved by reading the page tables directly. Then we allocated a large chunk of memory buffer in the spying enclave and determined their physical addresses. Using the reverse-engineering tool provided by the original DRAMA attack~\cite{pessl2016drama}, we picked two memory addresses $p$ and $p'$, as stated above. The attack is illustrated as in Figure~\ref{fig:drama}. $p$ and $p'$ are accessed in turns without any delay. The access latency of memory block $p$ is measured to determine whether the target address $d$ in the victim enclave has just been visited.

\textit{Second, obtain fine-grained timers in enclaves.}
An unexpected challenge in executing this attack is the lack of a reliable clock. The SGXv1 processor family does not provide timing information within an enclave: the instructions such as \texttt{RDTSC} and \texttt{RDTSCP} are not valid for enclave programs. To measure time, a straightforward way is making system calls, which is heavyweight, slow and inaccurate, due to the variation in the time for processing \texttt{EEXIT} and calls.  A much more lightweight solution we come up with utilizes the observation that an enclave process can access the memory outside without mode-switch.  Therefore we can reserve a memory buffer for smuggling CPU cycle counts into the attack enclave. Specifically, a thread outside the enclave continuously dumps the cycle counts to the buffer and the attack thread inside continuously reads from the buffer. Although the race condition between them brings in noise occasionally due to our avoidance of mutex for supporting timely interactions between the threads, most of the time we successfully observed a timing difference between a row hit and a row conflict when probing the target enclave addresses. We use this method to measure the access latency of $p$.

\begin{figure}
\includegraphics[width=\columnwidth]{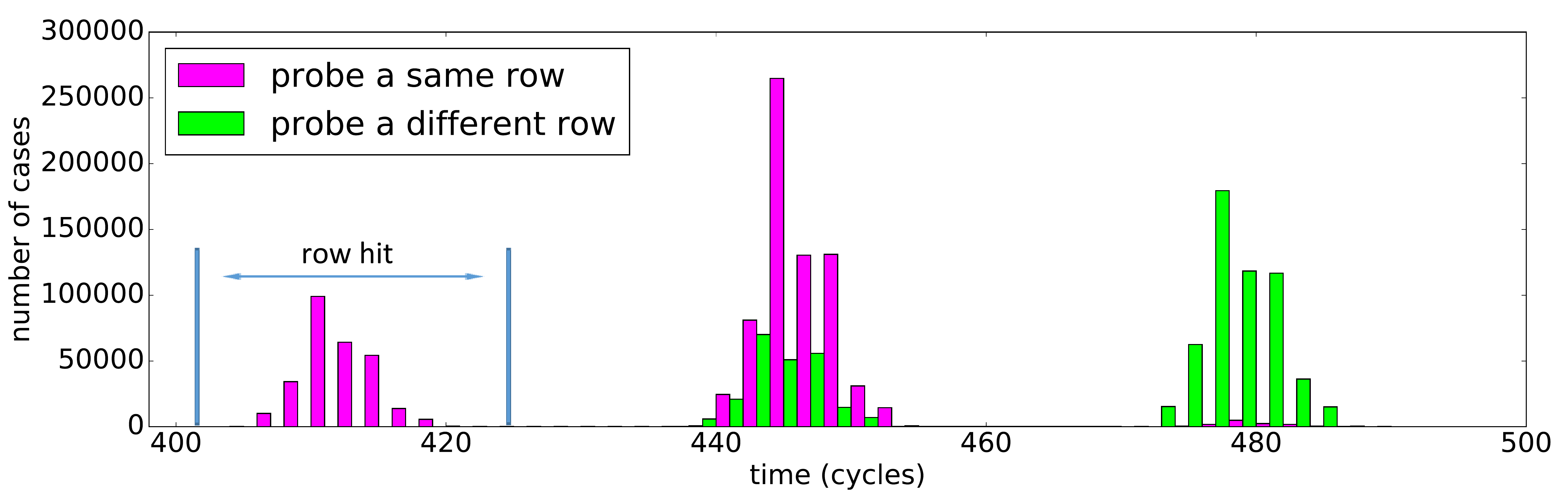}
\caption{Distribution of access latency for probing the same row and a different row.}
\label{fig:dramtiming5}
\end{figure}

\begin{figure}
\centering
\begin{lstlisting}
/*An input dependent branch from gap library*/
Obj SumInt(Obj opL, Obj opR) {
	// initialize temp variables
	// ...
	
	// adding two small integers
	if( ARE_INTOBJS( opL, opR) ) {
		if(SUM_INTOBJS(sum, opL, opR))
			return sum;
		cs = INT_INTOBJ(opL)+INT_INTOBJ(opR);
		// ...
	}
	// adding one large integer and small integer	
	else if( IS_INTOBJ(opL) || IS_INTOBJ(opR) ) {
		// ...
	}
	// add two large integers
	else {
		// ...
	}
}
\end{lstlisting}
\caption{An input-dependent branch in Gap 4.8.6.}\label{lst:gap}
\end{figure}

\vspace{3pt}\noindent\textbf{Evaluation.}
First we evaluate the accuracy of the timer we build for the attack. We designed a simple enclave process continuously visiting $d$ with \path{clflush} instruction forcing the memory accesses to reach a DRAM row. An evaluator enclave utilized the timer to measure the access latency of $p$ (the address on the same row as $d$), as well as the access latency of $p'$ (the address on a different row), 1 million times each. Figure~\ref{fig:dramtiming5} shows the distributions of the access latency measured by the evaluator enclave during these accesses. As we see here the cases of DRAM row hit can be easily identified based on the timing difference observed through our timer (the left-most part of its distribution).

As an example, we ported Gap 4.8.6 to Graphene-SGX, targeting an input-dependent branch which is illustrated in Figure~\ref{lst:gap}. Gap is a
software package implementing various algebra algorithms. It uses a non-integer
data type for values that cannot fit into 29 bits, otherwise the values are
stored as immediate integers. In our experiment we had the victim enclave
running the \texttt{SumInt} operation every 5 $\mu$s. We set the range for a row hit detection as within 400-426 cpu cycles. To further reduce false positives brought by prefetching, we disabled hardware prefetches on the victim core by updating \texttt{MSR} 0x1A4. With the cache-DRAM
attack targeting the instructions in line 8, our attack enclave could detect
whether the branch in line 7 was taken with a probability of 14.6\% and <1\%
false positive rate. Moreover, we only observed a 2\% slowdown of enclave
program in our experiment.

\vspace{3pt}\noindent\textbf{Discussion.}
The cache-DRAM attack achieves a
spatial accuracy of 64 byte which is as accurate as the \flushreload cache attacks. In the meanwhile it ensures that only the targeted cache set is primed which
further reduces the false positives caused by accesses of shared DRAM rows. The attack can be more powerful for a dedicated attacker by reserving a DRAM bank exclusively for the victim and spying enclaves.

%% file: tex/discussion.tex
\section{Mitigation and Discussion}
\label{sec:discussion}

\subsection{Analysis of Attack Surfaces}
\label{subsec:analysis}

\begin{table}
\centering \footnotesize{
\caption{Analysis of side-channel attack surfaces.}
\label{tab:analysis}
\begin{tabular}[t]{c|c|c|c}
\Xhline{1pt}
\textbf{Vectors} & \textbf{Accuracy} &
\textbf{AEX} & \textbf{Slowdown} \\
\Xhline{0.5pt}
\textbf{i/dCache \primeprobe} & 2MB &  high & high \\
\textbf{L2 Cache \primeprobe} & 128KB & high & high \\
\textbf{L3 Cache \primeprobe} & 16KB &  none & modest \\
\textbf{page faults} & 4KB & high & high \\
\textbf{B/T-SPM} & 4KB & modest & modest \\
\textbf{HT-SPM} & 4KB & none & modest \\
\textbf{cross-enclave DRAMA} & 1KB & none & high \\
\textbf{cache-DRAM} & 64B & none & minimal \\
\Xhline{1pt}
\end{tabular}} \\[4pt]
\begin{minipage}{0.98\columnwidth}
\centering
\end{minipage} \\[-10pt]
\end{table}

Table~\ref{tab:analysis} summarizes the characteristics of the memory
side-channel attacks discovered over different vectors.  The data here were
collected from the system configuration in Table~\ref{tab:configuration} and a
PRM size of 128MB.  The value under the \textit{Accuracy} column shows the
spatial accuracy of the attack vectors.  For example, the iCache \primeprobe
channel has an accuracy of 2MB (i.e., 128MB/64): that is, detecting one cache
miss in one of the iCache sets could probably mean any of 2MB of the physical
memory being accessed. The larger the number is, the coarser-grained the vector
will be. The attack with the finest granularity is the cache-DRAM attack, which
is 64 bytes, equivalent to the \flushreload cache attacks. However,
note that due to lack of shared memory pages---as EPC pages only belong to one
enclave at a time---\flushreload cache attacks are not feasible on SGX enclaves.
It is also worth noting that the calculation of the accuracy does not
consider knowledge of the physical memory exclusively used by the target
enclave. This information can help improve the granularity even further.
\primeprobe cache attacks on iCache, dCache and L2 cache induce high volume of AEXs.
This does not take HyperThreading into consideration. If so, both AEX numbers
and slowdowns will become \textit{modest}. Most of the attack vectors that need
to frequently preempt the enclave execution will induce \textit{high} overheads.
The cross-enclave DRAMA needs to disable cache to conduct effective attacks,
therefore inducing \textit{high} slowdown.  What is not shown in the table is
\textit{temporal observabilities}. Except for page-fault attacks, all other
attacks have temporal observabilities, as they allow observing finer-grained
information than allowed by their basic information unit, which are leaked
through timing signals.

\ignore{are deterministic, monitoring page accesses
whenever a new page is visited. Non-deterministic side channels suffer from
noise to different extent, missing observations at random intervals or having
several victim memory accesses batched together in the same observation.  For
example, \primeprobe attacks take certain time to probe all cache sets that need
to monitor, which could be slow, and if multiple memory accesses have been made
by the enclave program during one round of memory probing, the adversary is not
able to tell them apart. The SPM attack achieves a comparable temporal granularity with page-fault attacks, however it still suffers from missing oberservations if the inspection rate is low. Cross enclave DRAMA attack has a low temporal granularity. It could achieve much better temporal granularity if cache is disabled on the core of victim process so that the victim process runs orders of magnitude slower.}

\vspace{5pt}\noindent\textbf{Other attack vectors not listed}.
\flushreload cache attacks against cached PTE entries are one attack vector that
we have not listed in Table~\ref{tab:analysis}. As a PTE entry shares cache line with 7 more PTE entries, the spatial accuracy is 4KB$\times8$ = 32KB. The attack can achieve the spatial accuracy of 4KB if PTE entries are intentionally organized. Combining SPM and DRAMA attacks will also introduce a
new attack vector. We did not show these attacks due to the similarity to the ones we demonstrated.

\subsection{Effectiveness of Existing Defenses}
\label{subsec:defense}


\vspace{3pt}\noindent\textbf{Deterministic multiplexing}. Shinde
\etal~\cite{Shinde:2015:PYF} proposes a compiler-based approach to
opportunistically place all secret-dependent control flows and data flows into
the same pages, so that page-level attacks will not leak sensitive information.
However, this approach does not consider cache side channels or DRAM side
channels, leaving the defense vulnerable to cache attacks and DRAMA.

\vspace{5pt}\noindent\textbf{Hiding page faults with transactional memory}.
T-SGX~\cite{shih:tsgx} prevents information leakage about page faults inside enclaves by
encapsulating the program's execution inside hardware-supported memory
transactions. Page faults will cause transaction aborts, which will be handled by
abort handler inside the enclave first. The transaction abort handler will
notice the abnormal page fault and decide whether to forward the control flow to
the untrusted OS kernel. As such, the page fault handler can only see that the page fault happens on the page where the abort handler is located (via register \texttt{CR2}). The true faulting address is hidden.

However, T-SGX cannot prevent the \accessed enabled memory side-channel attacks.
According to Intel Software Developer's manual~\cite{IntelDevelopmentManual}, transaction
abort is not strictly enforced when the \accessed and \textit{dirty} flags of the
referenced page table entries are updated.\ignore{``Memory accesses within a
transactional region may require the processor to set the Accessed and Dirty
flags of the referenced page table entry. The behavior of how the processor
handles this is implementation specific. Some implementations may allow the
updates to these flags to become externally visible even if the transactional
region subsequently aborts. Some Intel TSX implementations may choose to abort
the transactional execution if these flags need to be updated.''} This means
there is no security guarantee that memory access inside transactional region is
not leaked through updates of the page table entries.

\vspace{5pt}\noindent\textbf{Secure processor design}.
Sanctum~\cite{Costan:2016:sanctum} is a new hardware design that aims to protect
against both last-level cache attacks and page-table based attacks. As Sanctum
enclave has its own page tables, page access patterns become invisible to the
malicious OS. Therefore, the page-faults attacks and SPM attacks will fail.
However, Sanctum does not prevent cross-enclave DRAMA attack. As a matter of
fact, Sanctum still relies on the OS to assign DRAM regions to enclaves, create
page table entries and copy code and data into the enclave during enclave
initialization.  Since OS knows the exact memory layout of the enclave, the
attacker can therefore run an attack process in a different DRAM region that
shares a same DRAM row as the target enclave address.

\vspace{5pt}\noindent\textbf{Timed execution}.
Chen \etal~\cite{Chen:2017:dejavu} proposes a compiler-based approach, called
\dejavu, to measure the execution time of an enclave program at the granularity
of basic blocks in a control-flow graph. Execution time larger than a threshold
indicates that the enclave code has been interrupted and AEX has occurred. The
intuition behind it is that execution time measured at the basic block level
will not suffer from the variations caused by different inputs. Due to the lack of
timing measurements in SGX v1 enclaves, \dejavu\ constructs a software clock
inside the enclave which is encapsulated inside Intel Transactional
Synchronization Extensions (TSX). Therefore, the clock itself will not be
interrupted without being detected.
It was shown that \dejavu\ can detect AEX with high fidelity.  Therefore, any of
the side-channel attack vectors that induce high volume of AEX will be detected
by \dejavu. However, those not involving AEX in the attacks, such as T-SPM or HT-SPM attacks
will bypass \dejavu\ completely.

\vspace{5pt}\noindent\textbf{Enclave Address Space Layout Randomization}. SGX-Shield~\cite{seo:sgx-shield} implemented fine-grained ASLR when an enclave program is loaded into the SGX memory. However the malicious OS could still learn the memory layout after observing memory access patterns in a long run as SGX-Shield does not support live re-randomization.

\subsection{Lessons Learned}
\label{subsec:lessions}

Our analysis of SGX memory side channels brings to light new attack surfaces and new capabilities the adversary has. Here are a few thoughts about how to mitigate such risks on the SGX platform, and more generically, for the emerging TEE.   

\vspace{5pt}\noindent\textbf{SGX application development.} Our research shows that the adversary can achieve fine-grained monitoring of an enclave process, through not only pages and cache channels, but also inter-page timing, cross-enclave DRAM and HyperThreading. It is important for the SGX developer to realize the impacts of these new attack surfaces, which is critical for building enclave applications to avoid leaks through the new channels. For example, she can no longer hide her secret by avoiding page-level access patterns, since intra- or inter-page timings can also disclose her sensitive information. 

\vspace{5pt}\noindent\textbf{Software-level protection.} Defense against SGX side-channel leaks can no longer rely on the assumptions we have today. Particularly, such attacks do not necessarily cause an anomalously high AEX rate. Blending sensitive information into the same memory pages is not effective against attacks with finer spatial granularity. Also, the adversary may also use a combination of multiple channels to conduct more powerful attacks.  

\vspace{5pt}\noindent\textbf{Hardware enhancement.} Most memory side channels we know so far can be mitigated through hardware changes, e.g., partitions of caches/DRAM and keeping enclave page tables inside EPC, etc. In some cases, such changes could be the best option. Further research is expected to better understand the issue and the impacts of the related side channels, making the case to Intel and other TEE manufacturers for providing hardware-level supports. 

\vspace{5pt}\noindent\textbf{Big picture.}  Over years, we observe that many side-channel studies follow a
similar pattern: a clever attack is discovered and then researchers immediately embark on the defense against the attack. However, in retrospection, most defense proposals fail to consider the bigger
picture behind the demonstrated attacks, thus they are unable to offer
effective protection against the adversary capable of quickly adjusting strategies, sometimes not even meaningfully raising the bar to the variations of the attacks.
The ongoing research on SGX apparently succumbs to the same pitfalls.  We hope that our study can serve as a new start point for rethinking the ongoing effort on SGX security, inspiring the follow-up efforts to better understand the fundamental limitations of this new TEE and the ways we can use it effectively and securely.

%% file: tex/relatedwork.tex
\section{Related Work}
\label{sec:relatedwork}

\vspace{5pt}\noindent{\textbf{Paging-based side channels}.}
It has been shown in previous studies that page-level memory access patterns can
leak secrets of enclave programs under a variety of
scenarios~\cite{Xu:2015:controlled, Shinde:2015:PYF}. This type of leakage is
enabled by enforcing page faults during enclave's execution, by marking selected
memory pages to be non-present or non-executable. As such, data accesses or
code execution in these pages will be trapped into the OS kernel, and
the malicious OS will learn which page is accessed by the enclave program.
Page-fault side-channel attack is one attack vector of the memory side-channel
attack surface we explore in this paper.

\revise{Concurrently and independently, Van \etal also propose paging-based attacks on SGX. They report two attacks: one exploits the updates of \accessed (Vector 4) and \dirtyflags (Vector 5) of the referenced PTEs, and the other is a \flushflush or \flushreload side-channel attack on the referenced PTEs (Vector 3).}

\revise{
\ignore{It is further shown that Vector 3 could enable a precise, instruction-granular enclave interrupt mechanism.}  
Although similar to our approach in terms of utilizing the \accessed to avoid page faults, the attack proposed has not been designed to be truly stealthy, minimizing interrupts produced when it is executed. Actually, it can introduce even \textit{more} AEXs, as demonstrated by their evaluation, rendering the attack less effective in the presence of existing protection, such as T-SGX~\cite{shih:tsgx} and \dejavu~\cite{Chen:2017:dejavu}. By comparison, our research reveals multiple avenues to reduce the interrupt frequencies, showing that a paging attack can be made stealthy when it is used together with timings or TLB flushing through HyperThreading, thwarting all existing defense. Further, our study also highlights other side-channel vectors in SGX memory management, providing evidence for the credible threats they pose (i.e., the Cache-DRAM attacks). 
}

\vspace{5pt}\noindent{\textbf{Branch prediction side channels}.}
A very recent study explores branch prediction units as side channels to infer sensitive control flows inside SGX enclaves~\cite{Lee:2016:SGXbranch}. The
memory side-channel attack surface does not include attack vectors through
branch prediction. Both are important to the understanding of side-channel
security of SGX.

\vspace{5pt}\noindent\textbf{Cache Side Channels}.
Cache side-channel attacks under the threat model we consider in this paper
(i.e., access driven attacks~\cite{Gullasch:2011:CGB}) have been demonstrated on
x86 architectures, including data caches (and also per-core L2 unified
caches)~\cite{Percival:2005:CMF, Osvik:2006:CAC, Neve:2006:AAC,Tromer:2010:ECA,
Gullasch:2011:CGB, Hund:2013:PTS}, instruction caches~\cite{Aciicmez:2007:YMA,
Aciicmez:2010:NRI, Zhang:2012:CSC}, and inclusive LLCs~\cite{Yarom:2014:FHR,
Yarom:2014:ROE, Benger:2014:JLB, Zhang:2014:CSA, Irazoqui:2014:WMF,
Yarom:2015:LLC, Irazoqui:2015:SCA, Gruss:2015:CTA, Oren:2015:SSP,
Inci:2015:seriously, Kayaalp:2016:HSA, Lipp:2016:ARMageddon, Zhang:2016:ROF}.
\revise{Some recent studies~\cite{brasser2017software,hahnel2017high,moghimi2017cachezoom,gotzfried2017cache} have shown that the above side channels are still feasible on SGX enclaves.}
We briefly confirmed the effectiveness of cache side-channel attacks in our paper, while the focus of this work is a broader attack surface than caches.

\vspace{5pt}\noindent{\textbf{SGX Side-Channel Defenses.}}
Most known defenses are designed specifically to page-fault side-channel attacks. Shinde \etal~\cite{Shinde:2015:PYF} proposed a compiler-based approach to transform cryptographic programs to hide page access patterns that may leak information. Shih \etal~\cite{shih:tsgx} proposed T-SGX which exploits Intel Transactional Synchronization Extensions (TSX) to prevent page faults from revealing the faulting address. Costan \etal~\cite{Costan:2016:sanctum} proposed a secure enclave architecture that is similar to SGX but resilient to both page-fault and cache side-channel attacks. Chen \etal~\cite{Chen:2017:dejavu} proposed \dejavu, a compiler-based approach to instrument enclave programs so that they can measure their own execution time between basic blocks in their control-flow graph. These research prototypes were designed without fully understanding the memory side-channel attack surface, thus fall short in offering effective protection against the attacks demonstrated in this work.

%% file: tex/conclusion.tex
\section{Conclusion}
\label{sec:conclusion}

We report the first in-depth study of SGX memory side channels in the paper.  Our study summarizes 8 attack vectors in memory management, ranging from TLB to DRAM. Further we demonstrate a set of novel attacks that exploit these channels, by leveraging \accessed, timing, HyperThreating and DRAM modules. Compared with the page-fault attack, the new attacks are found to be stealthier and much more lightweight, with effectiveness comparable with the known attack in some cases. Most importantly, our study broadens the scope of side-channel studies on SGX, reveals the gap between proposed defense and the design weaknesses of the system, and can provoke the further discussion on how to use the new TEE techniques effectively and securely.

%% file: tex/acknowledgement.tex
\section{Acknowledgement}
\label{sec:ack}
We are grateful to Taesoo Kim, the shepherd of our paper and the anonymous reviewers for their helpful comments. This work was supported in part by NSF 1408874, 1527141, 1566444 and 1618493, NIH 1R01HG007078, ARO W911NF1610127 and NSFC 61379139. Work at UIUC was supported in part by NSF CNS grants 12-23967, 13-30491 and 14-08944.